\documentclass[12pt]{article}    
\usepackage{graphics}
\begin{document}
\rm
\begin{quote} \raggedleft TAUP 2674-2001
\end{quote}
\begin{center}{\bf \large
The Effect of a Magnetic Flux Line in Quantum Theory}
\end{center}
\begin{center}{\bf Y. Aharonov$^{(a,b)}$, T. Kaufherr$^{(a)}$}\footnote{email
address trka@post.tau.ac.il}
\end{center}
(a) \emph{\small School of Physics and Astronomy, Tel Aviv
University,
Tel Aviv 69978, Israel}\\
(b)\emph {\small Department of Physics, University of South
Carolina,
 Columbia, S.C. 29208}\
\begin{center}{\bf Abstract}\
\end{center}\
\indent The nonloclal exchange of the conserved, gauge invariant
quantity\\
$e^{ \frac{i}{\hbar}
(p_{k}-\frac{e}{c}A_{k})L^{k}},\;L^{k}=const.,\;k=1,2$ between the
charged particle and the magnetic flux line (in the $k=3$
direction), is responsible for the Aharonov-Bohm effect. This
exchange occurs at a definite time, before the wavepackets are
brought together to interfere, and can be verified experimentally.
\vglue 0.1in \noindent PACS numbers: 03.65.Vf, 03.65.Ta, 03.65.Ud
\newpage\vglue 0.3333in \noindent {\bf I. Introduction} \vglue
0.33333in
\indent The Aharonov-Bohm (A-B) effect [1] is usually described in
terms of the shift in the interference pattern created by a
charged particle moving outside a solenoid. This description
leaves open the question regarding the time when the effect
occurs. Since the shift is due to a change in the relative phase
between the two wave packets of the particle, it seemed natural to
look for an answer in the time evolution of the relative phase.
But the relative phase is gauge dependent; it is only when the two
wave packets meet that it becomes  gauge invariant. Thus it seems
that quantum theory has contrived to eliminate any trace of when
the effect occurs, and that gauge invariance has some part in
this.\\\indent Below we show that the relevant variables are the
gauge invariant displacement operators $e^{  \frac{i}{\hbar}
(p_{k}-\frac{e}{c}A_{k})L^{k}},\;\;L^{k}=const., \;k=1,2$ for a
flux line in the $k=3$ direction (secII). In the A-B setup, these
otherwise conserved quantities are exchanged nonlocally and at a
definite time, between the charged particle and the flux line
(secsIII,IV). This accounts for the A-B effect (secsII,III). The
exchange alters the velocity distribution of the charged particle
and can therefore be verified experimentally (secIII).
\vglue 0.3333in \noindent {\bf II. Modular Momentum} \vglue
0.33333in
 \indent Quantum interference, unlike
classical interference$^{1}$, is a manifestation of an underlying
dynamics. To see this, consider a setup consisting of a charged
particle moving in the vicinity of a parallel, infinite plate
capacitor, but prevented from entering the electric field.
Consider first the position and momentum of the particle. The
Heisenberg equations of motion are,
\begin{eqnarray}
\dot{\vec{x}}&=&
\frac{i}{\hbar}\left[H,\vec{x}\right]=\frac{\vec{p}}{m}\;, \\
\dot{\vec{p}}&=& \frac{i}{\hbar}\left[H,\vec{p}\right]=e\vec{E}\;.
\end{eqnarray}
Since the electric field vanishes in the region where the particle
can be found, the average position and momentum of the particle
are unaffected by the electric field. Also the averages of  higher
moments of position and momentum remain unaffected. So far the
situation in quantum theory is similar to the classical one. But
consider the displacement operator
\begin{equation}
f(p_x)=e^{\frac {i}{\hbar}p_xL}\,. \label{eq:II-1}
\end{equation}
\begin{equation}
\frac{df}{dt}=\frac{i}{\hbar}\left[H,f\right]=
\frac{ie}{\hbar}(\int_{x}^{x+L}E\,dx' \;)\,f \,,
\end{equation}
where $x$ denotes the position of the particle. We have thus
discovered a quantity that changes. Compare this with the
classical analog of $(\!\!~\ref{eq:II-1})$,
\begin{equation}
f(p_x)=e^{i2\pi \frac {p_x}{p_0}}\,,
\end{equation} where
$p_0=\frac{h}{L}$. It satisfies
\begin{equation}
 \frac{df(p_{x})}{dt}= -\{
H,f(p_{x})\}= e E\; \frac{\partial f}{\partial
p_{x}}=i\frac{2\pi}{p_0}\;eEf \,,
\end{equation} where $\{\}$
denote Poisson brackets. It too does not change unless the
particle actually comes in contact with the field. The
displacement operator $f(p_x)$ is a function of modular momentum
$p_{x}(mod\;p_0)$ [2, 3] defined by
\begin{equation}
p_{x}(mod\;p_0)=p_{x}-Np_0 \,,
\end{equation}
 and so that its eigenvalues satisfy
\begin{equation}
0 \leq p_{x}(mod\;\, p_0) \leq p_0 \;.
\end{equation}
$N$ is an operator having integer eigenvalues. \ It is instructive
to consider the displacement operator also in the
Schr$\ddot{o}$dinger representation. Let the state of the particle
be
\begin{equation}
\Psi_{\alpha} =\Psi _{1}+e^{i\alpha }\Psi _{2}\,,
\end{equation}
 where $\Psi _{1}$ and $\Psi _{2}$ are
non-overlapping wavepackets, and $\Psi _{2}(x)=\Psi _{1}(x-L)\;.$
$\alpha$ represents the change in the wavefunction of the charged
particle due to its interaction with the capacitor. The average
values of $x$ and $p_x$ do not depend on $\alpha$ nor do averages
of any polynomial in $x$ or $p_{x}\;.$ Yet$^{2}$
\begin{eqnarray}
\left \langle e^{i2\pi\frac{p_x}{p_0}} \right \rangle &=&
\frac{1}{2} e^{i\alpha }\;.
\end{eqnarray}
The implication of this is that the modular momentum has changed.
compare this with the interference of classical waves. Consider as
an example two nonoverlapping wavepackets of classical
electromagnetic radiation. Introducing a relative phase between
the two will not change the total energy and momentum
\(\frac{1}{8\pi}\int\left(\mathcal{E}^{2}+B^{2}\right)dv\),
\(\frac{1}{4\pi c}\int\left(\vec{\mathcal{E}}\times
\vec{B}\right)dv\), of the field. This of course continues to be
true even when the two wavepackets come to overlap, and a shift in
the interference pattern due to the change in relative phase is
observed. When the electromagnetic field is quantized, the
operator \(e^{ \frac{i}{\hbar} \frac{1}{4\pi
c}\int\left(\vec{\mathcal{E}}\times \vec{B}\right)dv \cdot
\vec{L}}=e ^{\frac{i}{\hbar} \vec{P} \cdot \vec{L}   } \), where
$\vec{P}$ is the total momentum, acts as a displacement operator
on every state of the field. Thus, for a single photon in a
superposition of two  wavepackets displaced by $\vec{L}$, this
operator depends on the relative phase in exactly the same way as
the modular momentum considered above does.\\
Modular momentum has its own conservation law. Consider a
collision between two systems 1 and 2 such as the charged particle
and capacitor, and let
\begin{eqnarray}
\pi _1 &=& \cos (2\pi \frac {p_1}{p_0})  \\ \pi _2 &=& \cos (2\pi
\frac {p_2}{p_0}) \;,
\end{eqnarray}
where $p_1$, $p_2$ are the components of the momentum in any given
direction. Conservation of momentum in that direction, i.e.,
\begin{equation}
p_1+p_2=p'_1+p'_2
\end{equation}
then implies that also $C=\cos 2\pi \frac{p'_1+p'_2}{p_0}$ is
conserved, which leads to the conservation law (see fig.1)
\begin{equation}
1-C^2=(\pi'_1)^2+(\pi'_2)^2-2C\;\pi'_1\;\pi'_2 \,, \label{eq:II-2}
\end{equation}
which, in general, describes an ellipse. During a collision,
modular momentum is exchanged under the constraint
$(\!\!~\ref{eq:II-2})$. Obviously, $\Delta \pi _2 $ need not in
general be equal to $-\Delta \pi _1$ (where $ \Delta \pi _1=\pi
_1^{'}-\pi _1,\Delta \pi _2=\pi _2^{'}-\pi _2\,.$) Instead, the
systems translate from one point of the ellipse to another.\\
Finally, we show that a change in modular momentum, occurring
while the wavepackets are separated, manifests itself later as a
shift in the interference pattern. Consider the case where, after
the capacitor is closed, the two packets are allowed to meet.
Then,
\begin{equation}
\Psi_{\alpha}(x,0)=\Psi_{1}e^{ik_{0}x}+e^{i\alpha}\Psi_{2}e^{-ik_{0}x}
\,.
\end{equation}
In this case, the average of
\begin{equation}
O(0)=\cos[\frac{1}{\hbar}p_{x}(0)L+2 k_{0}x(0)] \,,
\end{equation}
which is a function of modular momentum, depends on $\alpha$. We
use the Heisenberg representation. Since $p_{x}(0)=p_{x}(t)\equiv
p_{x} $ and $x(0)=x(t)-\frac{p_{x}}{m}t$ are constants of the
motion,
\begin{equation}
\mbox{\ \ \ \ \ \ \ \ \ } =\cos\left[\frac{1}{\hbar}p_{x}L+2
k_{0}\left(x(t)-\frac{p_{x}}{m}t\right)\right]\equiv O(t) \,,
\end{equation}
 and
\begin{equation}
<O(t)>\equiv\frac{1 }{2}\cos \alpha \,.
\end{equation}
In particular, at $T=\frac{mL}{2\hbar k_{0}}\;$ $O$ turns into the
local function $O(T)=\cos [2k_{0}x(t)]$, which signals
interference. $<O(T)>=\frac{1 }{2}\cos \alpha\;$  is then a
sufficient condition for the shift of the interference pattern.
\vglue 0.3333in \noindent {\bf III. The effect of the vector
potential on the velocity distribution} \vglue 0.33333in
\indent With the aid of the gauge invariant version of modular
variables it is possible to determine the time when the A-B effect
occurs. Consider a charged particle prepared in a superposition of
two wavepackets
\begin{equation}
\Psi =\Psi _{1}+\Psi _{2} \,, \label{eq:III-1}
\end{equation}
where $\Psi _{2}(y)=\Psi _{1}(y-L),\; L\gg \Delta y,\;\Delta y $
is the spread of the packets. The particle passes by a thin
solenoid enclosing a flux $\Phi$, but sufficiently far from it, so
that its overlap with the solenoid is negligible. Thus it always
moves in the field free region surrounding the solenoid. We want
to show that the distribution of the velocity in the direction of
the line connecting the two wavepackets
 changes when this line crosses the solenoid provided the solenoid
lies in the interval between the packets (see fig. 2). Without
loss of generality we can assume that the line connecting the two
packets is in the $y$ direction. With the magnetic field within
the solenoid pointing in the +z direction, let us choose the gauge
\begin{eqnarray}
A_x=-\Phi \delta(x) \theta(y) ,\;\;\;\;  A_y\equiv0 ,
\end{eqnarray}
where $\theta (y)$ is the step function. We choose this gauge so
that before and after the particle passes by the solenoid
$\vec{p}=m\vec{v}$, since the vector potential vanishes in that
region. The \mbox{Fourier} transform of the velocity distribution
is then (we take $\hbar=1$)
\begin{eqnarray}
\int P_{r}(mv_{y}) e^{i mv_{y} L} dv_{y} = \int P_{r}(p_{y}) e^{i
p_{y}L}dp_{y} =<\Psi |e^{i p_{y}L}|\Psi>=\frac{1}{2}\;.
\end{eqnarray}
After $\Psi _{2}$ has crossed the potential line, the particle's
state is given by
\begin{equation}
\Psi '=\Psi _{1}  +e^{-i \alpha }\Psi _{2} \,,
\end{equation}
where $\alpha=\frac {e \Phi }{c \hbar }$. Consequently, the
Fourier transform of the velocity distribution has changed by
\begin{equation}
\delta \left < e^{i mv_{y} L} \right >= \frac {1}{2} \left ( e^{-i
\alpha }-1 \right ). \label{eq:III-2}
\end{equation}
 It is instructive to consider the situation in the Coulomb gauge
 where\mbox{$\vec{A}=\frac{\Phi}{2\pi r}\hat{e}_{\varphi}$}.
Comparing the value of the Fourier transform of the velocity
distribution shortly before the particle has passed by the
solenoid,
 with its value immediately afterwards, we find that it has changed
 by$^{3}$
\begin{equation}
\delta \left < e^{i m\vec{v} \cdot \vec{L}} \right > = \frac
{1}{2} \left ( e^{-i\frac{e}{c}\oint \vec{A} \cdot d\vec{l} }-1
\right ) = \frac {1}{2} \left ( e^{-i\frac{e}{c}\Phi}-1 \right
),
\end{equation}
in agreement with $(\!\!~\ref{eq:III-2})$,
having used Stokes's theorem and
\begin{equation}
 e^{i (\vec{p}-\frac{e}{c}\vec{A}) \cdot \vec{L}}=e^{i\frac{e}{c}\int_{\vec{r}+\vec{L}}^{\vec{r}}
  \vec{A} \cdot d\vec{l}} e^{i\vec{p} \cdot \vec{L}}\;\;. \label{eq:III-3}
\end{equation}
Thus, the change in the Fourier transform of the velocity
distribution occurs when the line connecting the wave packets
crosses the solenoid, even though in this gauge there is no sudden
change in the relative phase between the packets. To test this
experimentally, we suggest to measure, at a given time t, the
velocity distribution of an ensemble of N charged particles
prepared in a state $\Psi$ such as $(\!\!~\ref{eq:III-1})$, and to
calculate its Fourier transform. Repeating this procedure, using
different ensembles, at a succession of times during the motion of
the charged particle towards the flux line and past it, should
verify the existence of the effect.\\$(\!\!~\ref{eq:III-2})$
causes the shift in the interference pattern. This follows by
applying the argument from the end of sec II to$^{4}$
$O(t)=\cos[m\dot{y}L+2k_{0}(y-\dot{y}t)]$. Thus it is the gauge
invariant exchange, occurring while the charged particle passes by
the solenoid, and not any subsequent change in the relative phase
(which is not gauge invariant) that is responsible for the shift
in the interference pattern.
\vglue 0.3333in \noindent {\bf IV. The effect of the vector
potential on the angular velocity distribution} \vglue 0.33333in
\indent  The exchange of modular angular velocity  occurs on a
circle. To demonstrate this, we refer to the same setup as before,
i.e., a charged particle in a superposition of two wavepackets
moving towards a solenoid. For the sake of simplicity we shall
consider this case in the reference frame where the center of mass
of the charged particle is at rest. We define the coordinate
system thus: the origin coincides with the center of mass of the
charged particle, the $y$ axis passes through the centers of the
two wavepackets. The state of the charged particle is given by
\begin{equation}
\Psi=\Psi_1+\Psi_2\,,
\end{equation}
where $\Psi_2(\varphi,r)=\Psi_1(\varphi -\pi, r)$. $\Psi_1$ and
$\Psi_2$ are two nonoverlapping wave packets that are at rest, r
is the distance of their centers from the origin (see fig. 3). The
flux enclosed by the solenoid is $\Phi$, with the magnetic field
pointing in the +z direction. The solenoid moves towards the
particle in the $-x$ direction . The modular angular velocity
about the z-axis is given by
\begin{equation}
e^{iIv_{\varphi}\pi}=e^{i(p_{\varphi}-\frac{e}{c}rA_{\varphi})\pi}=e^{i\frac{e}{c}\int_{\varphi'=\varphi+\pi,r'\equiv
r
}^{\varphi}\vec{A}\cdot\vec{dl'}}e^{ip_{\varphi}}\,,
\end{equation}
having used relation $(\!\!~\ref{eq:III-3})$. $p_{\varphi}$ and
$I$ are, respectively, the angular momentum and moment of inertia
about the $z$ axis. The expectation value equals
\begin{equation}
\left< e^{iIv_{\varphi}\pi} \right>=1\rightarrow
\frac{1+e^{-i\alpha}}{2}\rightarrow \cos \alpha,
\end{equation}
where $\alpha=\frac{e\Phi}{\hbar \,c}$. As the solenoid moves
toward the charged particle and then away from it, the average
angular modular velocity changes abruptly twice, on the solenoid's
crossing the circle of radius $r$ centered at the origin. This,
too, is an experimentally verifiable result.\\\\
\vglue 0.3333in \noindent {\bf V. Discussion} \vglue 0.33333in
\indent The commonly accepted view is that, in order to measure
the A-B effect, two wavepackets moving on the two sides of a
solenoid have to be brought together to interfere. This also
epitomizes the topological nature of the effect. The picture
behind this is that of the relative phase being collected
gradually while the wavepackets travel through the nonsimply
connected region surrounding the solenoid, from which the
potentials can never be completely eliminated. We have shown that
this is not so, that the effect can be experimentally ascertained
to occur at an earlier stage, when the wavepackets are still
separated. The effect results from a nonlocal exchange of a
conserved, gauge invariant modular quantity between the charged
particle and the solenoid, which occurs at a definite time. It is
somewhat disturbing that this quantity contains information about
a region which the particle has never traversed$^{5}$. For
example, in the context of section III, in the gauge
$A_{x}\equiv0,$ $A_{y}=\Phi \theta (x) \delta(y)$. Note
nevertheless that the measurement of velocity necessarily brings
the particle to the region of that part of the  vector potential
that has not yet affected the phase of the wavefunction.  But we
also know that the measurement gives the distribution of the
velocities as it existed before the measurement.\\
This research was supported in part by Grant No.62/01 of the
Israel Science Foundation, by the NSF Grant No. NSF 0140377, and
by the ONR Grant No. N00014-00-0383. T. K. acknowledges the
support of the Grant-in-Aid for Specially Promoted Research, No.
10101001, by Monbusho, Japan. Special thanks to Prof. H. Yoshiki
of Kure University, Hiroshima, who led this grant.
\newpage
\noindent {\bf Footnotes } \vglue 0.33333in
\noindent 1. Below, we shall consider the interference of
classical electromagnetic waves and also its quantum counterpart.
\vglue 0.33333in \noindent 2. Note that because $\Psi _{\alpha}$
is a non-analytic function of $x\;,$
\begin{eqnarray}
\left \langle e^{i2\pi\frac{p_x}{p_0}} \right \rangle &=& \left
\langle \sum_{n} \frac{(2\pi i
)^n}{n!}\left(\frac{p_x}{p_0}\right)^n\right \rangle \neq
 \sum_{n} \frac{(2\pi i
)^n}{n!}\left \langle\left(\frac{p_x}{p_0}\right)^n\right \rangle
\;.
\end{eqnarray}
\vglue 0.33333in \noindent 3. In the Coulomb gauge,
\begin{eqnarray}
A_x=\frac{\Phi}{2\pi }\left(-\frac{y}{r^{2}}\right),\;\;\;\;
A_y=\frac{\Phi}{2\pi }\left(\frac{x}{r^{2}}\right).
 \label{eq:A-1}
\end{eqnarray}
At $x=-\varepsilon$, the state of the particle is
\begin{equation}
\Psi =\Psi _{1}+  e^{i\varphi _{0}}\Psi _{2} = \psi
(y-\frac{L}{2}) +e^{i\varphi _{0}}\psi
(y+\frac{L}{2})\;,\end{equation} where we assume that $\psi(y)$ is
centered about the origin.
\begin{equation}
\varphi _{0}=\frac{e}{c}\int
_{\frac{L}{2}}^{-\frac{L}{2}}A_{y}(y',x=-\varepsilon)dy'\;.
\label{eq:A-2}
\end{equation}
At $x=+\varepsilon$, the state of the particle is
\begin{equation}
\Psi '=\Psi _{1}e^{iS_{1}}+  e^{i\varphi _{0}}\Psi _{2}e^{iS_{2}}
\;,
\end{equation}
where
\begin{eqnarray}
S_{1}&=&\frac{e}{c}\int _{-\varepsilon}^{\varepsilon}A_{x}(x',y=
\frac{L}{2})dx'\;,\label{eq:A-3}\\
 S_{2}&=&\frac{e}{c}\int
_{-\varepsilon}^{\varepsilon}A_{x}(x',y=
-\frac{L}{2})dx'\stackrel{(\!\!~\ref{eq:A-1})}{=}-S_{1}\;\;.
\label{eq:A-4}
\end{eqnarray}
We have
\begin{equation} <\Psi|e^{imv_{y}L}|\Psi>=\frac{1}{2}\;, \end{equation}
and
\begin{eqnarray}
& &\mbox{  } <\Psi'|e^{imv_{y}L}|\Psi'>=\int dy
\Psi'^{*}(y)e^{-i\frac{e}{c}\int_{y}^{y+L}A_{y}(y',x=\varepsilon)dy'}e^{ip_{y}L}\Psi'(y)
\nonumber\\ &\approx&
e^{-i\varphi_{0}}e^{iS_{1}}e^{-iS_{2}}e^{-i\frac{e}{c}\int_{-\frac{L}{2}}^{\frac{L}{2}}A_{y}(y,x=\varepsilon)dy}
\int dy \psi^{*}(y+\frac{L}{2}) \psi(y+\frac{L}{2})
\nonumber\\
&=&\frac{1}{2}e^{-i\frac{e}{c}\oint\vec{A}\cdot d\vec{l}}
=\frac{1}{2}e^{-i\frac{e}{c}\Phi}\;,
\end{eqnarray}
having used $(\!\!~\ref{eq:III-3})$, $(\!\!~\ref{eq:A-2}),
(\!\!~\ref{eq:A-3})$, $(\!\!~\ref{eq:A-4})$ and Stokes's theorem.
We finally obtain,
\begin{equation}
\delta \left < e^{i m\vec{v} \cdot \vec{L}} \right >=
<\Psi'|e^{imv_{y}L}|\Psi'> -<\Psi|e^{imv_{y}L}|\Psi>=
\frac{1}{2}\left (e^{-i\frac{e}{c}\Phi}-1\right). \label{eq:A-9}
\end{equation}
\vglue 0.33333in \noindent 4.  Here expressed in the
Schr$\ddot{o}$dinger representation.
\vglue 0.33333in \noindent5. In this connection, see also [4].\\\\
\noindent {\bf References:}
\begin{itemize}
\item[{[1]}] Y. Aharonov and D. Bohm, Phys. Rev. 115 (1959) 485.
\item[{[2]}] Y. Aharonov, H. Pendelton and A. Petersen, Int. J. Th. Phys. 2 (1969) 213.
\item[{[3]}]  Y. Aharonov, Proc. Int. Sym.
Foundations of  Quantum Mechanics in the Light of New Technology,
Ed. S. Kamefuchi et. al. Phys. Soc. Japan,(1984) 10.
\item[{[4]}]M. V. Berry, Proc. R. Soc. Lond. A392, (1984) 45.
\end{itemize}
\newpage
\vglue 0.33333in \noindent {\bf Captions for Figures:} \vglue
0.33333in
\noindent Fig. 1. Conservation law of modular momentum
\begin{eqnarray}
\pi _1^{'} &=& \cos (2\pi \frac {p_1^{'}}{p_0})  \nonumber\\ \pi
_2^{'} &=& \cos (2\pi \frac {p_2^{'}}{p_0})  \label{eq:I31}
\end{eqnarray}
$p_0$ is a constant. $p_1^{'}$, $p_2^{'}$ are the components of
the momentum in any given direction after a collision. Due to a
collision, the system moves from one point of the ellipse to
another. For example, from a to b.
\newline\noindent
Fig. 2. Charged particle in a superposition of three wavepackets
moving in the x-direction. Only the distributions of the
velocities in the directions AB and AC will change when the
respective lines cross the solenoid, while the distribution of the
velocity in the direction BC does not change during the motion.
\newline\noindent
Fig. 3. View from the reference frame where the center of mass of
the charged particle is at rest: As the solenoid crosses the
circle line of radius $r$, the modular angular velocity about the
$z$ axis of the charged particle changes.
\newline\noindent
\includegraphics{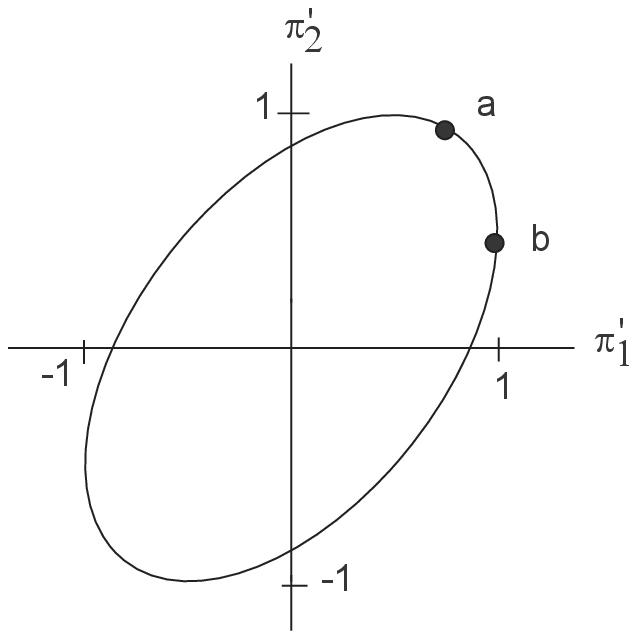}
\includegraphics{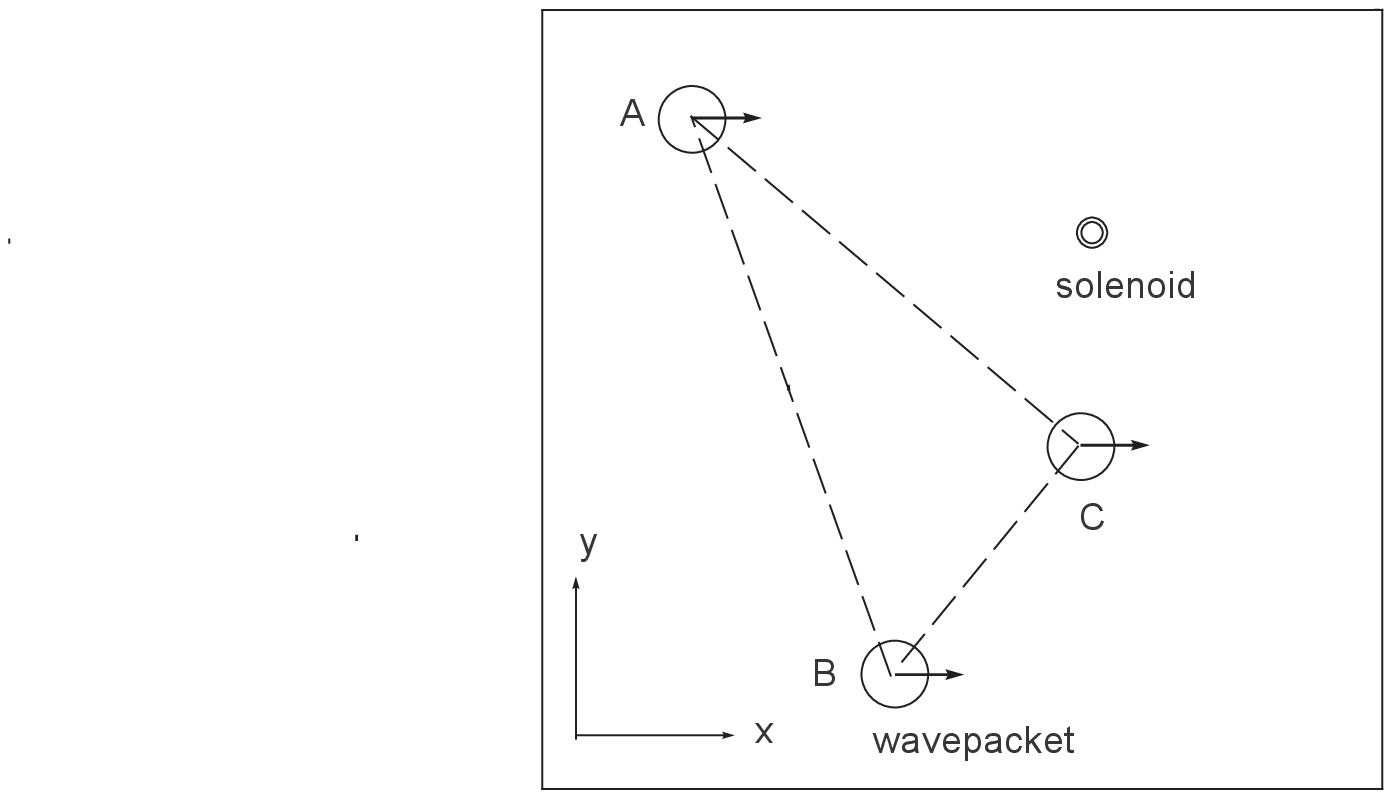}
\includegraphics{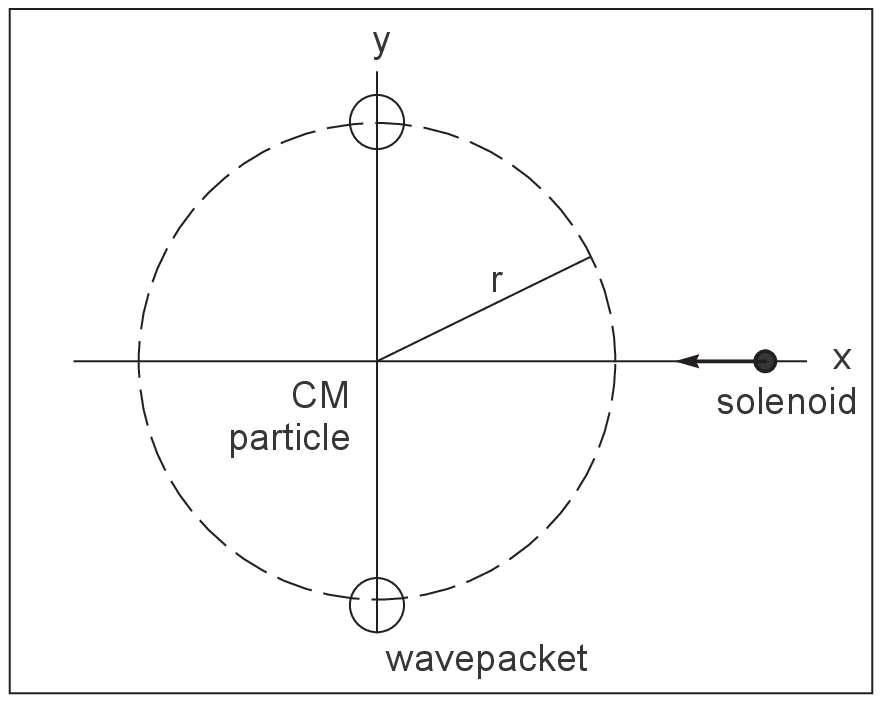}
\end{document}